\documentclass{webofc}
\usepackage[varg]{txfonts}   

\usepackage{graphicx}
\usepackage{xcolor}
\usepackage{caption}
\usepackage{subcaption}
\begin{document}
\makeatletter
\newsavebox{\@brx}
\newcommand{\llangle}[1][]{\savebox{\@brx}{\(\m@th{#1\langle}\)}%
  \mathopen{\copy\@brx\mkern2mu\kern-0.9\wd\@brx\usebox{\@brx}}}
\newcommand{\rrangle}[1][]{\savebox{\@brx}{\(\m@th{#1\rangle}\)}%
  \mathclose{\copy\@brx\mkern2mu\kern-0.9\wd\@brx\usebox{\@brx}}}
\makeatother

\title{The density of states method in Yang-Mills theories and first order phase transitions}
%
%

\author{\firstname{David} \lastname{Mason}\inst{1}\fnsep\thanks{\email{2036508@Swansea.ac.uk}} 
\and
\firstname{Biagio} \lastname{Lucini}\inst{2,3}\fnsep\thanks{\email{b.lucini@swansea.ac.uk}} 
\and
\firstname{Maurizio} \lastname{Piai}\inst{1}\fnsep\thanks{\email{m.piai@swansea.ac.uk}} 
\and
\firstname{Enrico} \lastname{Rinaldi}\inst{4,5,6,7}\fnsep\thanks{\email{erinaldi.work@gmail.com}} 
\and
\firstname{Davide} \lastname{Vadacchino}\inst{8}\fnsep\thanks{\email{davide.vadacchino@plymouth.ac.uk}}
}

\institute{Department of Physics, Faculty of Science and Engineering, Swansea University (Park Campus),
Singleton Park, SA2 8PP Swansea, Wales, United Kingdom
\and
Department of Mathematics, Faculty of Science and Engineering, Swansea University (Bay Campus),
Fabian Way, SA1 8EN Swansea, Wales, United Kingdom
\and
Swansea Academy of Advanced Computing, Swansea University (Bay Campus), Fabian Way, SA1 8EN Swansea, Wales, United Kingdom
\and
Physics Department, University of Michigan, Ann Arbor, MI 48109, USA
\and
Theoretical Quantum Physics Laboratory, Cluster of Pioneering Research, RIKEN, Wako, Saitama 351-0198, Japan
\and
Interdisciplinary Theoretical \& Mathematical Science Program (iTHEMS), RIKEN, Wako, Saitama, 351-0198, Japan
\and
Center for Quantum Computing (RQC), RIKEN, Wako, Saitama 351-0198, Japan
\and
Centre for Mathematical Sciences, University of Plymouth, Plymouth, PL4 8AA, United Kingdom
          }

\abstract{Extensions of the standard model that lead to first-order phase transitions in the early universe can produce a stochastic background of gravitational waves, which may be accessible to future detectors. Thermodynamic observables at the transition, such as the latent heat, can be determined by lattice simulations, and then used to predict the expected signatures in a given theory. In lattice calculations, the emergence of metastabilities in proximity of the phase transition may make the precise determination of these observables quite challenging,  and may lead to large uncontrolled numerical errors. In this contribution, we discuss as a prototype lattice calculation the first order deconfinement transition that arises in the strong SU(3) Yang-Mills sector.  We adopt  the novel logarithmic linear relaxation method, which can provide a determination of the density of states of the system with exponential error suppression. Thermodynamic observables can be reconstructed with a controlled error, providing a promising direction for accurate model predictions in the future.

}
\maketitle
\section{Introduction}
\label{intro}

Proposals for physics beyond the standard model (BSM) based on strongly interacting non-Abelian gauge sectors can provide an origin for a variety of phenomenologically interesting scenarios, such as dark matter and composite Higgs models (see, e.g., Refs.~\cite{Bennett:2017kga,Bennett:2019jzz,Bennett:2019cxd,Bennett:2022yfa,Kulkarni:2022bvh} 
for examples based on the Sp(4) gauge group). Our current understanding of these proposals is not yet detailed and precise, due to their strong-coupling nature. A signature with great potential discovery (or exclusion) reach is the generation of stochastic gravitational wave backgrounds from phase transitions in the early universe. In fact, many strongly interacting models are expected to undergo a change in behaviour between an early universe deconfined phase of quark gluon plasma and the late universe confined phase. Models characterised by a first-order phase transition lead to bubble nucleation which drives the generation of gravitational waves  (see, e.g., Refs.~\cite{huang2021testing,Halverson:2020xpg,Kang:2021epo}, and ~\cite{Reichert:2022naa} in these proceedings). The signature of the gravitational waves produced through bubble nucleation can in principle be predicted looking at thermodynamic observables measured at the phase transition. Of particular importance are the latent heat, which drives the expansion, and the surface tension, acting as a frictional term.

In principle, Lattice Field Theory is a prime tool to compute  non perturbatively observables related to the production of relic stochastic gravitational waves in such strongly-coupled models. However, the physical metastabilities that arise near criticality limit the power and reach of commonly used lattice algorithms, and call for the development of new ones. In this contribution, we discuss a novel application of the linear logarithmic relaxation (LLR) algorithm~\cite{langfeld2012density}, by analysing the thermodynamics of the phase transition in lattice simulations. This algorithm avoids the problems that affect standard Monte Carlo methods in a regions of metastability and hence enable us to calculate thermodynamic observables with robust error estimates. In the next section we discuss the general thermodynamics properties of non-Abelian lattice gauge theories at finite temperature, focusing in particular on the aforementioned algorithmic difficulties associated with metastability. In Sect.~\ref{sec:LLR} the LLR method is reviewed. Then, in Sect.~\ref{sec:results}, some preliminary results are discussed for the SU(3) pure gauge theory (see also Ref.~\cite{Borsanyi:2022xml} for a review and a recent high-precision calculation). Finally, Sect.~\ref{sect:conclusions} contains our summary. We remark that a similar approach is being undertaken 
for SU(4) (see Ref.~\cite{springer2021density} and the talk by F. Springer at this conference).

\section{First order phase transitions and metastable dynamics}
\label{sec:meta}

The Euclidean space-time is discretised onto an isotropic lattice with $\tilde{V}/a^4=N_t \times N_s^3$ sites, where $a$ is the lattice spacing. Since in this work we interpret the lattice field theory as a statistical mechanics system in four dimensions, for simplicity we set $a = 1$. Periodic boundary conditions are imposed in all directions. The spatial lattice size, $N_s$, is much larger than the temporal one, $N_t$. For fixed $N_t$, the temperature is set by changing the lattice spacing $a$ non-perturbatively, via  the lattice coupling $\beta$.  

The Wilson action for SU($N_c$) pure gauge theory is 
\label{sec:lattice}
\begin{equation}
    \label{eqn:lattice_action}
 S =  \sum_{j=0}^{\tilde{V}} \sum_{\mu; \nu > \mu}\left(1 - \frac{1}{N_c} \Re(\textrm{Tr}[U_{\mu\nu}(j)])\right) \,,
\end{equation}
where $U_{\mu\nu}(j)$ denotes a plaquette, which is the parallel transport of the lattice links $U_{\mu}(i) \in$ SU($N_c$) around the elementary square of the lattice in the $\mu \nu$ plane, starting at lattice site $j$.

In standard importance sampling approaches, configurations are  generated via Monte Carlo Markov chain methods, with a combination of heat bath and over-relaxation steps, with a probability distribution $P_{\beta}(E)$ for a value of the action $S = E$ given by
\begin{equation}
    P_\beta (E) = \frac{1}{Z(\beta)} e^{- \beta E} \ , \qquad Z(\beta) = \int \left( {\cal D} U_\mu(i) \right) e^{- \beta S} \ , 
\end{equation}
where the integral is taken over all SU($N_c$)-valued link variables $U_\mu(i)$. 
Vacuum expectation values (VEV) of observables, $\langle O \rangle$, are estimated as (ensemble) averages  over the sampled configurations.

When focusing on the thermodynamics at the phase transition, the observables of interest are the average plaquette, which we denote as $u_p$, the specific heat (or plaquette susceptibility), the Polyakov loop, and the Polyakov loop susceptibility. The Polyakov loop is defined in terms of parallel transports of the link variables  on paths of fixed spatial coordinates that wrap around the temporal direction. The Polyakov loop VEV~\cite{mclerran1981quark} is the order parameter of the spontaneous breaking of the global centre symmetry associated with the deconfinement phase transition. It is given by 
\begin{equation}
    \label{eqn:polyakov loop}
\left\langle  l_p  \right\rangle _\beta = \left\langle \frac{1}{N_c N_s^3} \sum_{\vec{n}_s} \textrm{Tr}\left(\prod_{n_t = 0}^{N_t - 1} U_0(n_t, \vec{n_s})\right)  \right\rangle_\beta  \ \  \begin{cases} = 0 \text{ confined phase} \\
\neq 0 \text{ deconfined phase}
\end{cases} \ ,
\end{equation}
where we have introduced the notation $i = (n_t, \vec{n}_s)$, with $n_t$ being the temporal and $\vec{n}_s$ the spatial coordinates of the lattice site $i$.

In SU($N_c$) Yang-Mills gauge theory with $N_c > 2$, the deconfinement phase transition is of first order and therefore exhibits metastable dynamics around the critical point. To accurately measure the thermodynamic quantities of interest for the system with a lattice calculation based on importance sampling, the configuration space must be  sampled  by an ergodic Markov chain. To ensure ergodicity, in the metastable region the Markov chain must allowed to tunnel between the two phases several times. But standard local Monte Carlo methods only consider small changes in each link,  suppressing drastic changes. Consequently, these algorithms struggle to overcome the potential barrier separating the two phases. While for small lattices, and small number of colours $N_c$  (between three and six),  this is not a problem, in the limit of large volumes (or large $N_c$) the potential barrier grows. More configurations are therefore required to ensure the system can explore all the configuration space, with the associated computational time growing exponentially with the volume, to compensate for the tunneling probability being exponentially suppressed with the volume.

\section{Linear logarithmic relaxation method}
\label{sec:LLR}
The linear logarithmic relaxation (LLR) method allows, in principle, to avoid the problems in simulating the equilibrium distribution of the metastable dynamics near the critical region of  first-order phase transitions. It uses Monte Carlo algorithms with the purpose of recovering microcanonical information, which is then used to reconstruct physical observables.

 In the LLR approach to a general statistical system the energy range of interest is broken down into intervals, with energy 
  $E_n -\delta_E/2 < E < E_n + \delta_E/2$. The interval size $\delta_E$ should be chosen small enough for a Taylor expansion of physical quantities in $\delta_E$ to be a good approximation in each interval. The total action of Yang-Mills theories
   is used as an analogue for the energy\footnote{We use the term energy to express the value of the action on a particular configuration or set of configurations.}, $E=6\tilde{V}(1 - u_p)$. For each interval the density of states,
\begin{equation}
\rho(E)=\int [ D \phi ] \delta(S[\phi]-E) \ ,
\end{equation}
is approximated numerically. 
The partition function of the statistical system with Hamiltonian $S$
(or, equivalently, path integral of the field theory with action $S$) is
rewritten in terms of $\rho(E)$, and becomes a simple one-dimensional integral over the allowed energy interval:
\begin{equation}
\label{eq:partition_function}
Z(\beta)=\int [ D \phi ] e^{-\beta S [ \phi ]} = \int d E \rho(E) e^{-\beta E} \,.
\end{equation}
Similarly, the VEV computed at coupling $\beta$ of an observable $O(E)$ that strictly depends on the energy becomes 
\begin{equation}\label{eq:vev_obs}
 \langle O(E) \rangle_\beta=\frac{1}{Z(\beta)}\int dE \rho(E) O(E) e^{-\beta E}\ . 
\end{equation}

We approximate $\rho(E)$ in a small interval of amplitude $\delta_E$ around $E_n$, by Taylor expanding its logarithm, $\ln(\rho(E))\approx a_n(E-E_n) + c_n$, where $c_n$ is a constant. By rearranging this equation and using textbook statistical mechanics considerations,  the Taylor coefficients $a_n$ are identified with the inverse microcanonical temperature at energy $E_n$, as $a_n = (d\ln\rho/ dE)|_{E=E_n} = (ds / dE)|_{E=E_n} \equiv 1 / t_n$, where $s$ is the entropy.

The individual Taylor expansions, repeated over a sequence of intervals $E_0 < ... < E_{n} < E_{n+1} < ... < E_N$, for $E_{n+1} = E_n + \delta_E$, can be combined by imposing continuity, to obtain,  for the density of states in an interval $E_n-\delta_E/2 < E < E_n+\delta_E/2$, the expression
\begin{equation}
    \label{eq:RM_rho}
    \rho (E) \approx  \rho_0 \exp\left(\sum_{k=0}^{n-1}(a_k \delta_E) + a_n (E-E_n + \delta_E / 2)\right)\,.
\end{equation}
This expression can be evaluated for any E in the range $E_0 - \delta_E/2 < E < E_N + \delta_E/2$. 
The overall factor $\rho_0$ can be fixed arbitrarily, as it drops in computations of ensemble averages.

\begin{figure}
\sidecaption
\centering
\includegraphics[width=0.6\textwidth]{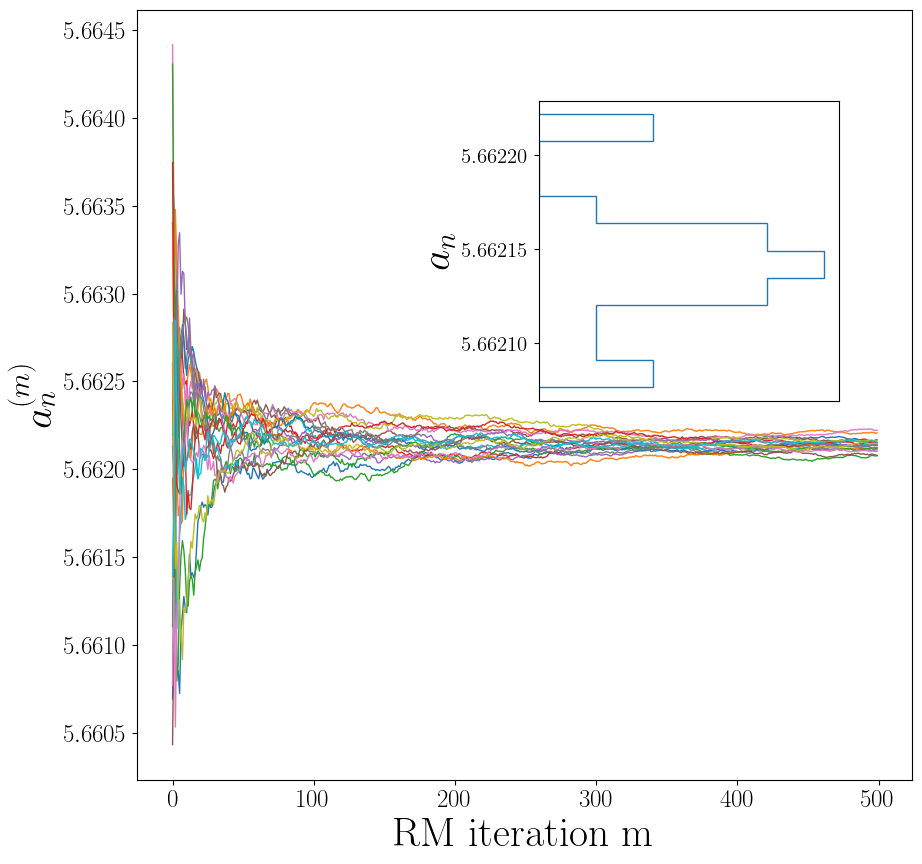}
\caption{An sample of 20 RM trajectories for SU(3) pure gauge theory on a $4 \times 20^3$ lattice with periodic boundary conditions in all directions and an energy interval of size $\delta_E/6\tilde{V} = 0.000374281$ centred at $E_n/6\tilde{V}= 0.459324241$. The inset shows the obtained $a_n^{(m)}$ distribution at the truncation value $m = 500$. 
}
\label{fig:RM}     
\end{figure}

\section{Results}
\label{sec:results}

We can think of the coefficients $a_n$ in terms of the temperature that gives rise to an energy distribution that is symmetric around the centre of the interval $E_n$. This implies $\llangle \Delta E \rrangle_{a_n} \equiv \llangle E - E_n \rrangle_{a_n} = 0$, where the double brackets denote the sampled mean over configurations restricted to the interval.
To solve this equation, we follow Ref.~\cite{langfeld2016efficient} and use the Robbins-Monro (RM) approach~\cite{RobbinsMonro:1951}. We choose an initial guess for $a_n^{(0)} \approx a_n$. Then, using the equation $a_n^{(m+1)} = a_n^{(m)} - 12\llangle \Delta E \rrangle_{a_n^{(m)}} / \delta_E^2 (m+1)$, we  iteratively improve our guess of the true value of $a_n$. For a suitable initial value, after a transient, $a_n^{(m)}$ will oscillate around the root $a_n$ with an increasingly suppressed amplitude, and $\lim_{m\to \infty} a_n^{(m)} = a_n$.  An example of the calculation of $a_n$ for the SU(3) system considered here is shown in Fig.~\ref{fig:RM}. 
We assess the size of the truncation error of the iteration by repeating the stochastic calculation 20 times, beginning with the same initial guess, and averaging  the results. As $m$ increases, all the repeats appear to converge towards the same final value. The values of $a_n$ have been calculated for intervals $E_n$ that cover the energy range relevant for the physical problem we are interested in. To avoid ergodicity problems associated with the restricted sampling, umbrella sampling was used when applying RM iterations, as described in Ref.~\cite{lucini2016overcoming}.

The VEV of observables that strictly depend on the energy at a coupling $\beta$ are computed by plugging the numerically determined values of $a_n$ into Eq.~(\ref{eq:vev_obs}), as follows
\begin{equation}\label{eq:vev_obs_2}
 \langle O(E) \rangle_\beta= \frac{1}{Z(\beta)} \sum_{n=0}^{N}\int_{E_n-\frac{\delta_E}{2}}^{E_n + \frac{\delta_E}{2}}dE \ \rho_0 \exp\left(\sum_{k=0}^{n-1}(a_k \delta_E) + a_n (E-E_n + \delta_E / 2)\right) O(E)e^{-\beta E} \,. 
\end{equation}
An analogous expression for $Z(\beta)$ is left implicit. This equation displays a sum over contributions from all intervals,
but notice the suppression of contributions for which the associated inverse micro-canonical temperature is far from 
the coupling $\beta$ we are calculating at.

\begin{figure}
     \centering
     \begin{subfigure}[b]{0.49\textwidth}
         \centering
         \includegraphics[width=\textwidth]{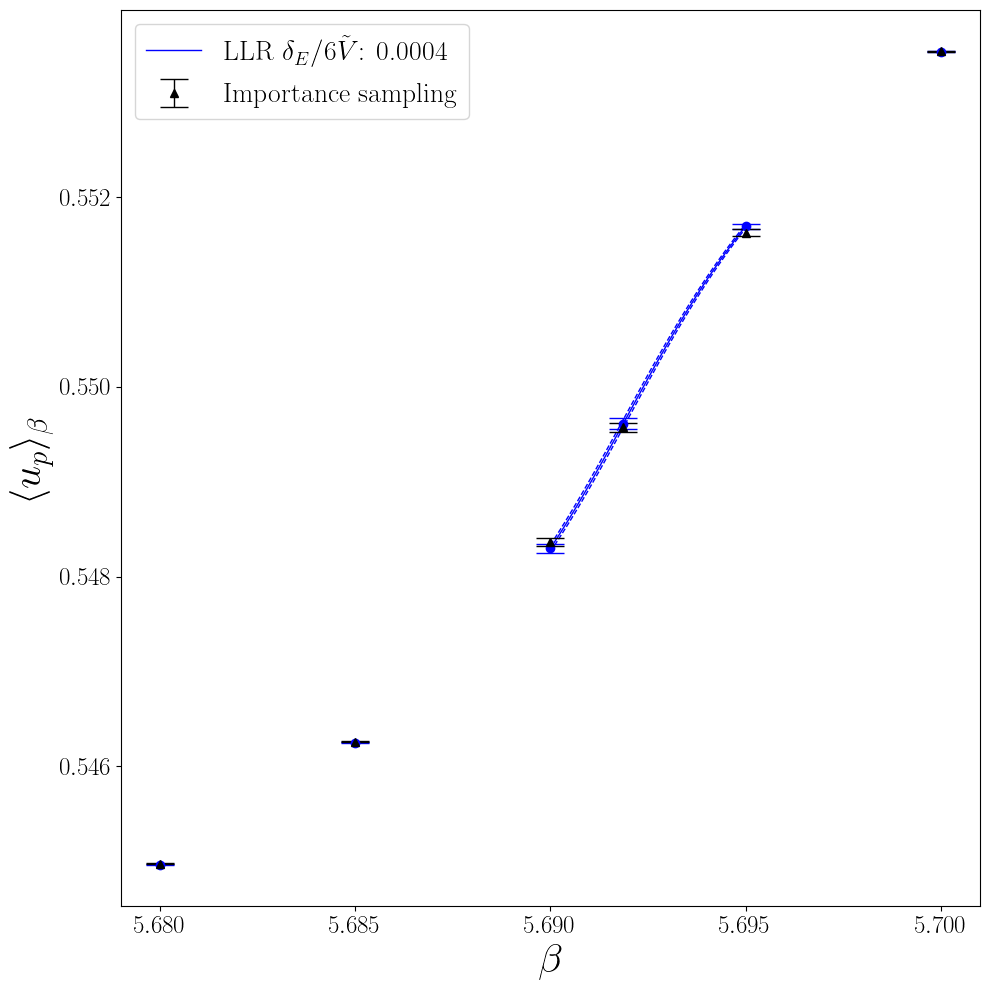}
         \label{fig:plaq}
     \end{subfigure}
     \hfill
     \begin{subfigure}[b]{0.49\textwidth}
         \centering
         \includegraphics[width=\textwidth]{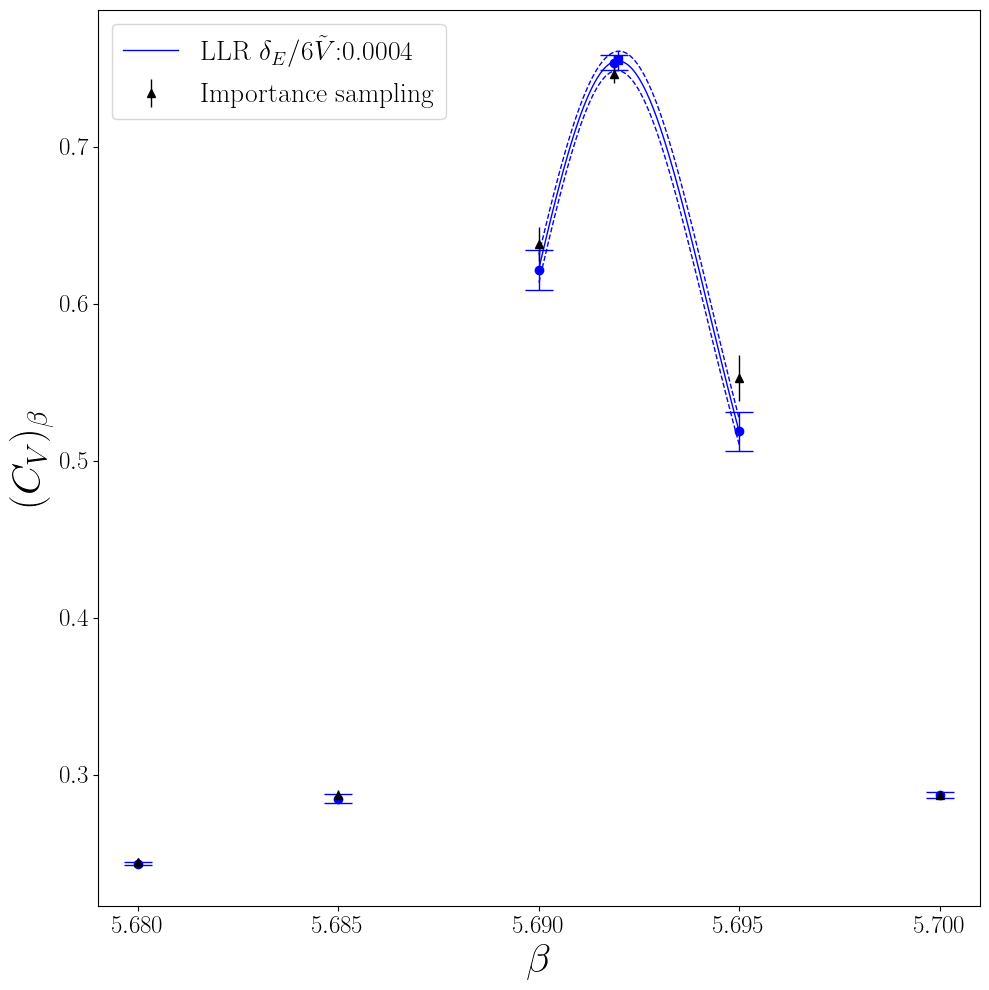}
         \label{fig:Cv}
     \end{subfigure}
        \caption{Thermodynamic observables measured with the LLR method (blue circles), compared to results from standard importance sampling (black triangles), for SU(3) Yang-Mills gauge theory on a $4 \times 20^3$ lattice. The LLR calculation uses energy intervals of size $\delta_E/6\tilde{V} = 0.000374281$ in the energy range from $E_0/6\tilde{V} = 0.439487341$ to $E_{54}/6\tilde{V} = 0.459698522$. The blue curves are reconstructed observables from LLR method with a finer resolution in $\beta$, restricted to the region around the phase transition. Left panel:  average plaquette $\langle u_p \rangle$ against the coupling $\beta$. Right panel: specific heat $C_V \equiv \langle u_p^2 \rangle - \langle u_p \rangle^2$ against the coupling $\beta$.}
        \label{fig:Reconstructed Observables}
\end{figure}

Using Eq.~(\ref{eq:vev_obs_2}) we have calculated the VEV of the average plaquette, $\langle u_p \rangle_\beta$, and the specific heat, $C_V \equiv \langle u_p^2 \rangle_\beta - \langle u_p\rangle_\beta^2$. These are shown in Fig.~\ref{fig:Reconstructed Observables} plotted against the coupling value they are calculated at. To demonstrate the validity of this method, the plots include the values measured using standard importance sampling methods.   

\begin{figure*}
\sidecaption
\centering
\includegraphics[width=0.6\textwidth]{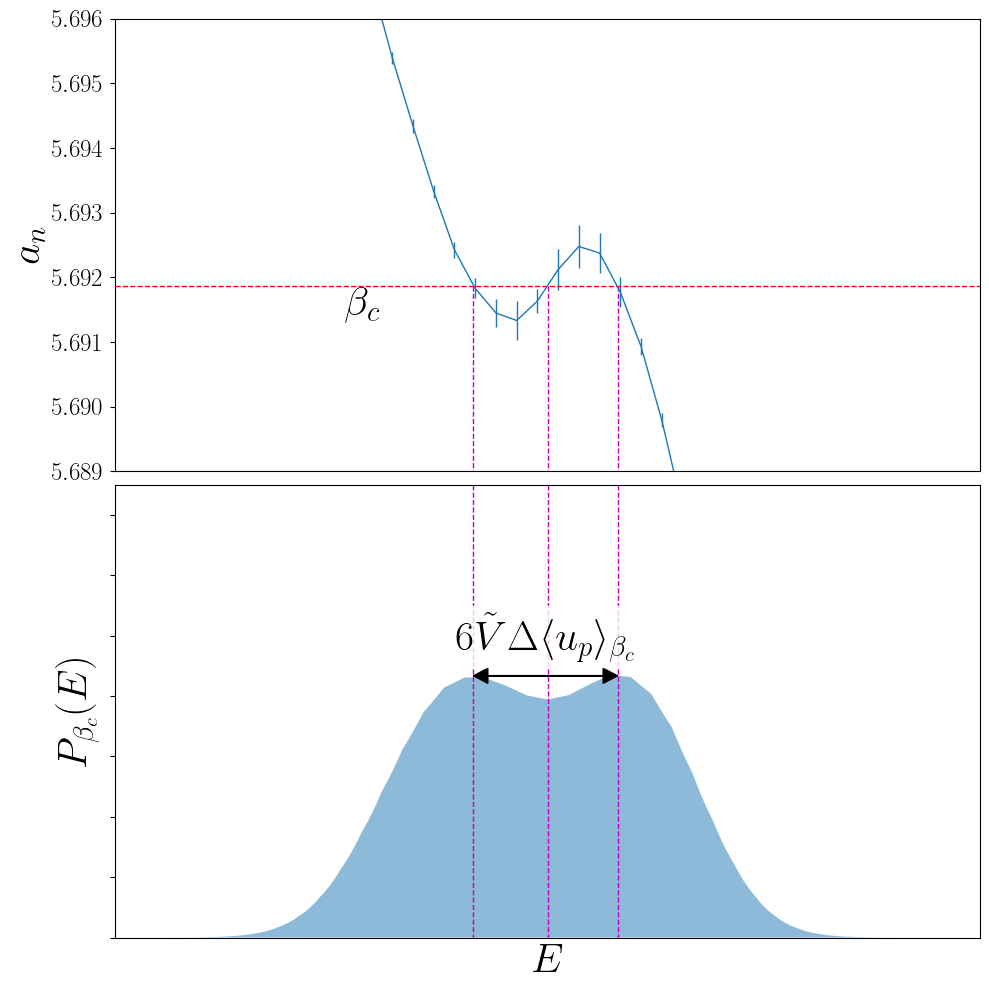}
\caption{Results of the LLR analysis of SU(3) Yang-Mills gauge theory on a $4 \times 20^3$ lattice with energy intervals of size $\delta_E/6\tilde{V} = 0.000374281$ in the energy range from $E_0/6\tilde{V} = 0.439487341$ to $E_{54}/6\tilde{V} = 0.459698522$.
The top panel shows the values of the micro-canonical inverse temperatures $a_n$ against the centre of the energy interval $E_n$, with a linear interpolation between the points.
The bottom panel shows the reconstructed probability distribution $P_{\beta_c}(E)$ of the energy $E$ at the critical coupling $\beta_c$.
The horizontal dashed line shows the location of the critical coupling, and the vertical lines are the average plaquette values at which $a_n = \beta_c$, which correspond to the locations of the extrema of the probability distribution.}
\label{fig:potential}       
\end{figure*}

An important observable for gravitational wave physics is the latent heat, $L_h$, of the transition. As shown, e.g., in Ref.~\cite{lucini2005properties}, the key thermodynamic  quantity that enters the calculation of $L_h$ is the average plaquette jump at criticality, $\Delta\langle u_p \rangle_{\beta_c}$. For the definition of the critical point, we take the value of $\beta$ at which both phases are equally probable. The corresponding energy distribution can be approximated with a double Gaussian in which the two peaks corresponding to the two different phases are of equal height~\cite{Challa:1986sk}. The critical point is found by taking the value of $\beta$ at which the specific heat has a maximum, as an initial guess, and then using a bisection algorithm to refine the estimate until a predefined tolerated different height between the peaks is met.  $\Delta\langle u_p \rangle_{\beta_c}$ is then the energy difference between the position of the confined peak and the position of the deconfined peak at the determined value of $\beta_c$ divided by $6\tilde{V}$, as shown in Fig.~\ref{fig:potential}.

The presence of metastability can be inferred directly from the non invertibility of  $a_n(E_n)$, as shown in Fig.~\ref{fig:potential}.The points at which $a_n = \beta_c$ correspond to the locations of the extrema of the distribution. The thermodynamics of the system can be further analysed using the micro-canonical ensemble. In analogy with a statistical mechanics system, we can define the temperature $t$, entropy $s$, and the free energy of the micro-states as follows:
\begin{equation}
    \label{eq:therm}
    t_k \equiv\frac{1}{a_k}, \qquad s \equiv \ln(\rho), \qquad F \equiv E - t s.
\end{equation}
 Similarly to the case of a Van der Waals gas below criticality, we expect the free energy to display a swallow tail structure (see, e.g., Ref.~\cite{Kubiznak:2012wp}), that indicates a first order transition. 
 
\begin{figure}
\sidecaption
\centering
\includegraphics[width=0.6\textwidth]{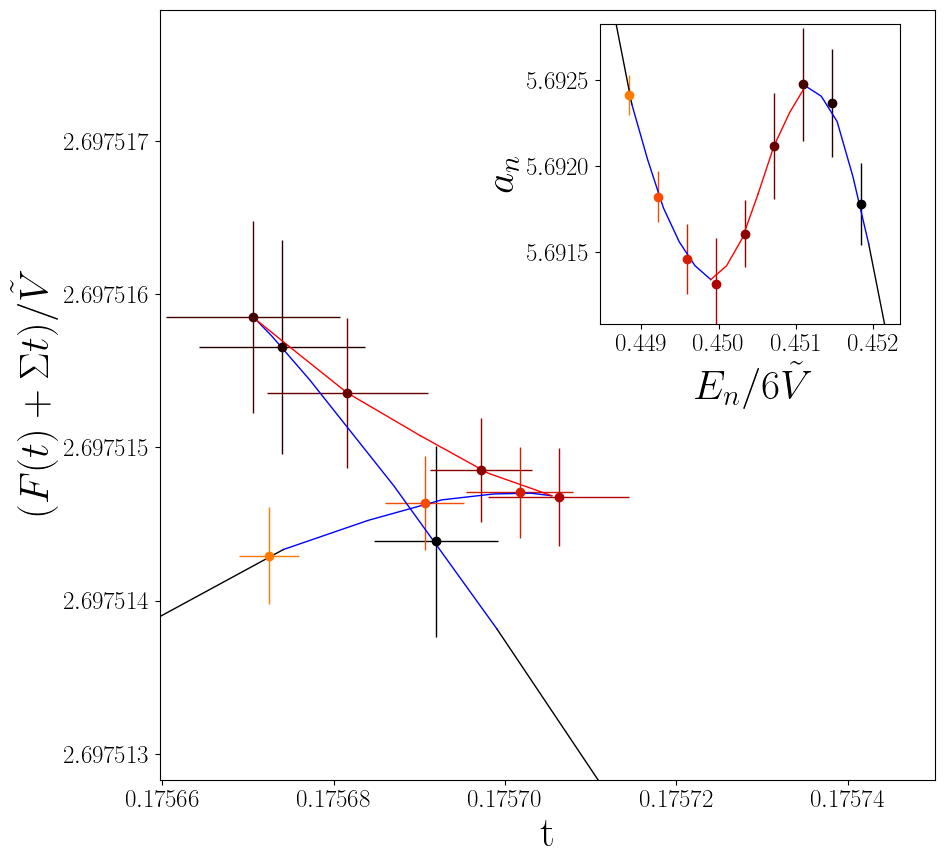}
\caption{The free-energy for SU(3) Yang-Mills gauge theory on a $4\times20^3$ lattice, computed in the LLR method with energy intervals of size $\delta_E/6\tilde{V} = 0.000374281$, in the energy range from $E_0/6\tilde{V} = 0.439487341$ to $E_{54}/6\tilde{V} = 0.459698522$. The free-energy is given by $F=E-t s$, where $E=6\tilde{V}(1-u_p)$ is the internal energy,  $s=\ln\rho$ is the entropy and $t =1/a_n$ is the temperature. $\Sigma$ is a constant, computed as the temperature average of $s - \ln\rho_0$ where $\rho_0$ originates in the density of states, Eq.~(\ref{eq:RM_rho}). The inset shows $a_n$ and $E_n$, for the corresponding points.}
\label{fig:freeenergy}       
\end{figure}

A subtlety emerges because the definition of the entropy depends on the constant $\rho_0$  in the density of states. $\rho_0$ contributes a term linear in the temperature to the free energy. As we are only interested in seeing the qualitative behavior of the free energy, we subtract a linear term $\Sigma t$, where $\Sigma$ is computed as the temperature average of the entropy within the energy range considered. The result is shown in Fig.~\ref{fig:freeenergy}. This clearly shows the telltale signs of a first order transition. Away from the critical region the free energy is single valued. In the critical region the free energy has three values at any given $t$. The highest such value is associated with the unstable vacuum, the middle one with the false vacuum and the lowest with the true vacuum. The critical point corresponds to the location at which the lines from the true and false vacua meet, resulting in a non-analyticity of the minimum of the free energy as a function of $t$.

\section{Conclusions}
\label{sect:conclusions}
First-order phase transitions in the early universe can give rise to a stochastic background of gravitational waves that can be used to constrain new physics. For strongly coupled new physics models, the relevant observables can in principle be measured using lattice field theory. With the aim to obtain robust results, we have explored the possibility of applying the LLR method, whereby the density of states of the system is determined numerically in the microcanonical ensemble and then used to reconstruct observables performing one-dimensional numerical integrals. We have tested the method for a SU(3) gauge theory close to criticality, on a single lattice volume and at a relatively coarse lattice spacing. The method has been validated using conventional lattice calculations, which are still viable for the chosen lattice parameters. Our approach has also shown clear advantages in terms of the quantities it can give us access to, which include the free energy, a precisely determined probability distribution of the states at criticality and the dependency of the microcanonical temperature on the energy. Those quantities, which cannot be robustly determined in importance sampling approaches, show clear singularities that carry direct information about the phase transition. The material presented here is the first step of a programme that aims to use the LLR method to compute more accurately and efficiently thermodynamic quantities in non-Abelian gauge theories.\\

~\\
\begin{acknowledgement}
{\bf Acknowledgements - } We would like to thank David Schaich and Felix Springer for discussions.
The work of D.~V. is partly supported by the Simons Foundation under the program ``Targeted Grants to Institutes'' awarded to the Hamilton Mathematics Institute.
The work of D.~M. is supported by a studentship awarded by the Data Intensive Centre for Doctoral Training, which is funded by the STFC grant ST/P006779/1.  
E.~R. was supported by Nippon Telegraph and Telephone Corporation (NTT) Research.
The work of B.~L. and M.~P. has been supported in part by the STFC 
 Consolidated Grants No. ST/P00055X/1 and No. ST/T000813/1. B.L. and M.P. received funding from the European Research Council (ERC) under the European Union’s Horizon 2020 research and innovation program under Grant Agreement No.~813942. The work of BL is further supported in part by the Royal Society Wolfson Research Merit Award WM170010 and by the Leverhulme Trust Research Fellowship No. RF-2020-4619. Numerical simulations have been performed on the Swansea SUNBIRD cluster (part of the Supercomputing Wales project) and AccelerateAI A100 GPU system, and on the DiRAC Data Intensive service at Leicester. The Swansea SUNBIRD system and AccelerateAI are part funded by the European Regional Development Fund (ERDF) via Welsh Government. The DiRAC Data Intensive service at Leicester is operated by the University of Leicester IT Services, which forms part of the STFC DiRAC HPC Facility (www.dirac.ac.uk). The DiRAC Data Intensive service equipment at Leicester was funded by BEIS capital funding via STFC capital grants ST/K000373/1 and ST/R002363/1 and STFC DiRAC Operations grant ST/R001014/1. DiRAC is part of the National e-Infrastructure.\\
{\bf Open Access Statement - } For the purpose of open access, the authors have applied a Creative Commons 
Attribution (CC BY) licence  to any Author Accepted Manuscript version arising.
\end{acknowledgement}

%
\bibliography{references}

\begin{thebibliography}{19}

\bibitem{Bennett:2017kga}
E.~Bennett, D.K. Hong, J.W. Lee, C.J.D. Lin, B.~Lucini, M.~Piai, D.~Vadacchino,
  JHEP \textbf{03}, 185 (2018), \texttt{1712.04220}

\bibitem{Bennett:2019jzz}
E.~Bennett, D.K. Hong, J.W. Lee, C.J.D. Lin, B.~Lucini, M.~Piai, D.~Vadacchino,
  JHEP \textbf{12}, 053 (2019), \texttt{1909.12662}

\bibitem{Bennett:2019cxd}
E.~Bennett, D.K. Hong, J.W. Lee, C.J.D. Lin, B.~Lucini, M.~Mesiti, M.~Piai,
  J.~Rantaharju, D.~Vadacchino, Phys. Rev. D \textbf{101}, 074516 (2020),
  \texttt{1912.06505}

\bibitem{Bennett:2022yfa}
E.~Bennett, D.K. Hong, H.~Hsiao, J.W. Lee, C.J.D. Lin, B.~Lucini, M.~Mesiti,
  M.~Piai, D.~Vadacchino, Phys. Rev. D \textbf{106}, 014501 (2022),
  \texttt{2202.05516}

\bibitem{Kulkarni:2022bvh}
S.~Kulkarni, A.~Maas, S.~Mee, M.~Nikolic, J.~Pradler, F.~Zierler (2022),
  \texttt{2202.05191}

\bibitem{huang2021testing}
W.C. Huang, M.~Reichert, F.~Sannino, Z.W. Wang, Physical Review D \textbf{104},
  035005 (2021)

\bibitem{Halverson:2020xpg}
J.~Halverson, C.~Long, A.~Maiti, B.~Nelson, G.~Salinas, JHEP \textbf{05}, 154
  (2021), \texttt{2012.04071}

\bibitem{Kang:2021epo}
Z.~Kang, J.~Zhu, S.~Matsuzaki, JHEP \textbf{09}, 060 (2021),
  \texttt{2101.03795}

\bibitem{Reichert:2022naa}
M.~Reichert, Z.W. Wang, \emph{{Gravitational Waves from dark composite
  dynamics}} (2022), \texttt{2211.08877}

\bibitem{langfeld2012density}
K.~Langfeld, B.~Lucini, A.~Rago, Physical Review Letters \textbf{109}, 111601
  (2012)

\bibitem{Borsanyi:2022xml}
S.~Borsanyi, K.~R., Z.~Fodor, D.A. Godzieba, P.~Parotto, D.~Sexty, Phys. Rev. D
  \textbf{105}, 074513 (2022), \texttt{2202.05234}

\bibitem{springer2021density}
F.~Springer, D.~Schaich, PoS \textbf{LATTICE2021}, 043 (2022),
  \texttt{2112.11868}

\bibitem{mclerran1981quark}
L.D. McLerran, B.~Svetitsky, Physical Review D \textbf{24}, 450 (1981)

\bibitem{langfeld2016efficient}
K.~Langfeld, B.~Lucini, R.~Pellegrini, A.~Rago, The European Physical Journal C
  \textbf{76}, 1 (2016)

\bibitem{RobbinsMonro:1951}
H.~Robbins, S.~Monro, Annals of Mathematical Statistics \textbf{22}, 400 (1951)

\bibitem{lucini2016overcoming}
B.~Lucini, W.~Fall, K.~Langfeld, PoS \textbf{LATTICE2016}, 275 (2016),
  \texttt{1611.00019}

\bibitem{lucini2005properties}
B.~Lucini, M.~Teper, U.~Wenger, Journal of High Energy Physics \textbf{2005},
  033 (2005)

\bibitem{Challa:1986sk}
M.S.S. Challa, D.P. Landau, K.~Binder, Phys. Rev. B \textbf{34}, 1841 (1986)

\bibitem{Kubiznak:2012wp}
D.~Kubiznak, R.B. Mann, JHEP \textbf{07}, 033 (2012), \texttt{1205.0559}

\end{thebibliography}

\end{document}